# QUANTUM SECURE DIRECT COMMUNICATION USING ENTANGLEMENT AND SUPER DENSE CODING

Ola M. Hegazy, Ayman M. Bahaa Eldin and Yasser H. Dakroury

*Computer and Systems Eng. Department, Ain Shams University, Cairo, Egypt*
ola_hegazy@yahoo.com, ayman.bahaa@eng.asu.edu.eg , ydakroury@mcit.gov.eg

Abstract: This paper introduces a new quantum protocol for secure direct communication. This protocol is based on Entanglement and Super-Dense coding. In this paper we present some basic definitions of entanglement in quantum mechanics, present how to use the maximally entangled states known as Bell States, and super dense coding technique to achieve secure direct message communication. Finally, we will apply some error models that could affect the transmission of the quantum data on the quantum channels, and how to treat these errors and acquire a safe transmission of the data.

## 1 INTRODUCTION

The aim of cryptography is to ensure that a secret message is transmitted between two users in a way that any eavesdropper cannot read it. Since classical cryptography relies on difficulty and infeasibility of computation to find the plain text, it is losing security more and more as computational power is increasing by technical innovations. In classical cryptography, it is generally accepted that one-time pad, which utilizes a previously shared secret key to encrypt the message transmitted in the public channel, is the only cryptosystem with proved security. Fortunately, quantum key distribution (QKD) (Bennett, 1984), the approach using quantum mechanics principles for distribution of secret key, can overcome this obstacle skillfully. Since both (QKD) and one-time pad have been proved secure (Lee, 2005), the cryptosystem of "QKD & one-time pad" is a perfect one when the security is concerned.

  Previously proposed QKDPs are the theoretical design (Bennett, 1984), security proof (Massey, 1988), and physical implementation (Bennett, 1992).

  Quantum secure direct communication (QSDC) (Boström, 2002, Deng, 2008) is another branch of quantum cryptography. Different from QKD, QSDC allows the sender to transmit directly the secret message (not a random key) to the receiver in a deterministic and secure manner. If it is designed carefully, a QSDC protocol can also attain unconditional security (Deng, 2003).

  The main objective of our research is to introduce a new protocol that guarantees more security of the transmission than the QKD and also saves more time, cost and gives more efficiency for the transmission, as it is using the super dense coding technique that transmit two classical bits by sending one quantum bit. In our protocol of the quantum secure direct communication we use the maximally entangled Bell states to encode the message bits on the basis of the super dense coding theorem, and then transmitting them on two quantum channels to the other side with less probability of the eavesdropping, and with no need for a pre-shared key that in turn needs many rounds to distribute, and also a public discussions to verify the correctness of the key.

## 2 BACKGROUND

The most important and interesting characteristics of the quantum mechanics is that the quantum state could not be measured without disturbing and changing the state of the particles (photons). So the use of quantum phenomenon will help in overcoming one of the most important eavesdropping problems; that is measuring the information without being discovered, so any attempt of Eve to measure the data during transmission will be known to Alice and Bob.

Also, another interesting feature of the quantum phenomenon is that any arbitrary quantum state cannot be cloned or copied and that is known as No-Cloning theorem (Nielsen, 2000). Of course, this will help in overcoming another eavesdropping problem which is copying the transmitted signal, so Eve cannot take a copy of the message during transmission. These two characteristics of the quantum phenomenon make it a stronger mechanism in securing the transmission path more than the classical transmission.

Quantum mechanics violates everyday intuition not only because the measured data can only be predicted probabilistically but also because of a quantum specific correlation called entanglement. Entanglement can be used to cause non local phenomenon. States possessing such correlations are called entangled states. Among these states, the states with the highest degree of entanglement (correlation) are called maximally entangled states or EPR states, as historically, the idea of a non local effect due to entanglement was pointed out by Einstein, Podolsky, and Rosen (Hayashi, 2006).

The pure quantum nature of entanglement is the property of non-local correlations between widely separated particles which have interacted in the past. To make particles entangled, it is necessary for them to interact at a point. In other words, the non-local property of entanglement is arisen from the local property of interaction (Lee, 2005).

## 3 THE SUPER DENSE CODING PROCEDURE

The super dense coding is a simple example of the application of quantum entanglement communication. The goal of this procedure is to transmit two classical bits by sending one quantum bit (qubit), so increasing the efficiency of the transmission.

Before starting the transmission, it is assumed that a third party has generated an entangled state, one of the Bell entangled state, for example $\left|\phi^+\right\rangle = \frac{1}{\sqrt{2}}\left(\left|00\right\rangle + \left|11\right\rangle\right)$, and then sends one of the two pairs of the entangled qubits to the sender 'Alice' and the other to the receiver 'Bob'.

When starting the transmission, Alice could send the single qubit in her possession to Bob after operating on it in such a way to encode two bits of the classical information to Bob.

As there are four possible values of the two classical bits Alice wishes to send to Bob: 00, 01, 10 and 11, then if Alice wants to send the two bits '00', she does nothing to her qubit just simply send it as it is. If she wants to send '10', she applies the phase flip Z to her qubit. If she wants to send '01', she applies the quantum NOT gate X, to her qubit. If she wants to send '11', she applies the iY gate to her qubit. The four quantum gates that are used here are: the Pauli matrices I, Z, X, iY, and combinations of them are applied as the U unitary operation that Alice performs on her half of the EPR pairs according to the diagram in fig.(1) (Benenti, 2004).

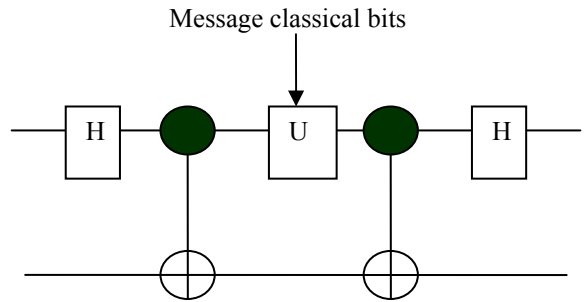

Figure 1: A quantum circuit implementing the super dense coding.

The use of these four different transformations results in the four states of Bell states as in the following equations:

$$00 \rightarrow \left|\varphi^+\right\rangle = \frac{1}{\sqrt{2}}\left(\left|00\right\rangle + \left|11\right\rangle\right) \quad (1)$$

$$10 \rightarrow \left|\varphi^-\right\rangle = \frac{1}{\sqrt{2}}\left(\left|00\right\rangle - \left|11\right\rangle\right) \quad (2)$$

$$01 \rightarrow \left|\psi^+\right\rangle = \frac{1}{\sqrt{2}}\left(\left|01\right\rangle + \left|10\right\rangle\right) \quad (3)$$

$$11 \rightarrow \left|\psi^-\right\rangle = \frac{1}{\sqrt{2}}\left(\left|01\right\rangle - \left|10\right\rangle\right) \quad (4)$$

Of course, the kind of the transformation of the operator $U$ will change according to the state that is generated in the first half of the circuit before the $U$ operator box, as we took an example of the state $\left|\phi^+\right\rangle$, but it could be any other Bell state. In all cases the generated state after the operator $U$ (the encoding circuit) will be also any other one of the Bell states but in different order according to the classical bits that will be sent.

## 4 THE NEW PROPOSED QSDC PROTOCOL USING SUPER-DENSE CODING

Our new protocol of the quantum secure direct communication uses the maximally entangled Bell states to encode the message bits using super dense coding that was mentioned above, and then transmit them on two quantum channels to the other side with less probability of the eavesdropping and more efficiency by sending two classical bits using one quantum bit (qubit).

### 4.1 Basic Idea of the Protocol

Using the idea of the super dense coding theorem, the message bits are input to the operator U selector, then according to their value (one of the four possible values 00, 01, 10 or 11), and also according to the state of the entangled pair that will be generated randomly (one of the Bell states) due to the inputs of the quantum selector $(|i_1\rangle, |i_0\rangle)$, the inputs to the $U$ Pauli operator will drive it to perform one of the four unitary operations mentioned above, the $U$ Pauli operators could be $I, X, iY$ or $Z$ and their matrix representations are as follows:

$$I = \begin{bmatrix} 1 & 0 \\ 0 & 1 \end{bmatrix}, X = \begin{bmatrix} 0 & 1 \\ 1 & 0 \end{bmatrix}, Y = \begin{bmatrix} 0 & -i \\ i & 0 \end{bmatrix}, Z = \begin{bmatrix} 1 & 0 \\ 0 & -1 \end{bmatrix}$$

These steps are applied using the block diagram of figure 2.

### 4.2 Assumptions of the Protocol

Alice will be the one who will produce the EPR pairs (Bell states carriers) in her side not a third party. Then she keeps one half for herself (and apply the encoding of operator U on it) and sends the other half to Bob.

The Bell states that will be generated by Alice will be chosen on random basis using a random generator, so we can get one of the different four maximally entangled Bell states, $|\phi^+\rangle, |\psi^+\rangle, |\phi^-\rangle,$ and $|\psi^-\rangle$.

### 4.3 Steps of the Protocol

1) Alice inputs the message bits (M), 2-bits by 2-bits, as one input to select the U operator as mentioned above with the other input that is one of the Bell states which is randomly generated.
2) Then after applying the unitary operator on the qubit of Alice, the result that is the output of the Alice encoding circuit $|\psi^o\rangle$ will be sent on one of the public quantum channel and on a spatially separated quantum channel the Bob's qubit (the half of Bell state).
3) So according to the different generated Bell states we will have the function matrix of U operator according to the following table and analysis:

a) In case of the carrier Bell state $|\phi^+\rangle$:

| 2-bit element | $U$ | $|\psi^o\rangle$ |
|---|---|---|
| 00 | $I$ | $(I \otimes I)|\phi^+\rangle = |\phi^+\rangle$ |
| 01 | $X$ | $(X \otimes I)|\phi^+\rangle = |\psi^+\rangle$ |
| 10 | $Z$ | $(Z \otimes I)|\phi^+\rangle = |\phi^-\rangle$ |
| 11 | $iY$ | $(iY \otimes I)|\phi^+\rangle = |\psi^-\rangle$ |

According to the following analysis:

$$(I \otimes I)|\varphi^+\rangle = \begin{bmatrix} 1 & 0 & 0 & 0 \\ 0 & 1 & 0 & 0 \\ 0 & 0 & 1 & 0 \\ 0 & 0 & 0 & 1 \end{bmatrix} \bullet \frac{1}{\sqrt{2}} \begin{bmatrix} 1 \\ 0 \\ 0 \\ 1 \end{bmatrix}$$
$$= \frac{1}{\sqrt{2}} \begin{bmatrix} 1 \\ 0 \\ 0 \\ 1 \end{bmatrix} = |\varphi^+\rangle \quad (5)$$

$$(X \otimes I)|\varphi^+\rangle = \begin{bmatrix} 0 & 0 & 1 & 0 \\ 0 & 0 & 0 & 1 \\ 1 & 0 & 0 & 0 \\ 0 & 1 & 0 & 0 \end{bmatrix} \bullet \frac{1}{\sqrt{2}} \begin{bmatrix} 1 \\ 0 \\ 0 \\ 1 \end{bmatrix}$$
$$= \frac{1}{\sqrt{2}} \begin{bmatrix} 0 \\ 1 \\ 1 \\ 0 \end{bmatrix} = |\psi^+\rangle \quad (6)$$

$$(Z \otimes I)|\varphi^+\rangle = \begin{bmatrix} 1 & 0 & 0 & 0 \\ 0 & 1 & 0 & 0 \\ 0 & 0 & -1 & 0 \\ 0 & 0 & 0 & -1 \end{bmatrix} \bullet \frac{1}{\sqrt{2}} \begin{bmatrix} 1 \\ 0 \\ 0 \\ 1 \end{bmatrix}$$
$$= \frac{1}{\sqrt{2}} \begin{bmatrix} 1 \\ 0 \\ 0 \\ -1 \end{bmatrix} = |\varphi^-\rangle \quad (7)$$

$$(iY \otimes I)|\varphi^+\rangle = \begin{bmatrix} 0 & 0 & 1 & 0 \\ 0 & 0 & 0 & 1 \\ -1 & 0 & 0 & 0 \\ 0 & -1 & 0 & 0 \end{bmatrix} \bullet \frac{1}{\sqrt{2}} \begin{bmatrix} 1 \\ 0 \\ 0 \\ 1 \end{bmatrix}$$
$$= \frac{1}{\sqrt{2}} \begin{bmatrix} 0 \\ 1 \\ -1 \\ 0 \end{bmatrix} = |\psi^-\rangle \quad (8)$$

b) In case of the carrier Bell state $|\phi^-\rangle$:

| 2-bit element | U | $|\psi^o\rangle$ |
|---|---|---|
| 00 | Z | $(Z \otimes I)|\phi^-\rangle = |\phi^+\rangle$ |
| 01 | XZ | $(XZ \otimes I)|\phi^-\rangle = |\psi^+\rangle$ |
| 10 | I | $(I \otimes I)|\phi^-\rangle = |\phi^-\rangle$ |
| 11 | iYZ | $(iYZ \otimes I)|\phi^-\rangle = |\psi^-\rangle$ |

According to the following analysis:

$$(Z \otimes I)|\varphi^-\rangle = \begin{bmatrix} 1 & 0 & 0 & 0 \\ 0 & 1 & 0 & 0 \\ 0 & 0 & -1 & 0 \\ 0 & 0 & 0 & -1 \end{bmatrix} \cdot \frac{1}{\sqrt{2}} \begin{bmatrix} 1 \\ 0 \\ 0 \\ -1 \end{bmatrix}$$
$$= \frac{1}{\sqrt{2}} \begin{bmatrix} 1 \\ 0 \\ 0 \\ 1 \end{bmatrix} = |\varphi^+\rangle \quad (9)$$

$$(XZ \otimes I)|\varphi^-\rangle = \begin{bmatrix} 0 & 0 & -1 & 0 \\ 0 & 0 & 0 & -1 \\ 1 & 0 & 0 & 0 \\ 0 & 1 & 0 & 0 \end{bmatrix} \cdot \frac{1}{\sqrt{2}} \begin{bmatrix} 1 \\ 0 \\ 0 \\ -1 \end{bmatrix}$$
$$= \frac{1}{\sqrt{2}} \begin{bmatrix} 0 \\ 1 \\ 1 \\ 0 \end{bmatrix} = |\psi^+\rangle \quad (10)$$

$$(I \otimes I)|\varphi^-\rangle = \begin{bmatrix} 1 & 0 & 0 & 0 \\ 0 & 1 & 0 & 0 \\ 0 & 0 & 1 & 0 \\ 0 & 0 & 0 & 1 \end{bmatrix} \cdot \frac{1}{\sqrt{2}} \begin{bmatrix} 1 \\ 0 \\ 0 \\ -1 \end{bmatrix}$$
$$= \frac{1}{\sqrt{2}} \begin{bmatrix} 1 \\ 0 \\ 0 \\ -1 \end{bmatrix} = |\varphi^-\rangle \quad (11)$$

$$(iYZ \otimes I)|\varphi^-\rangle = \begin{bmatrix} 0 & 0 & -1 & 0 \\ 0 & 0 & 0 & -1 \\ -1 & 0 & 0 & 0 \\ 0 & -1 & 0 & 0 \end{bmatrix} \cdot \frac{1}{\sqrt{2}} \begin{bmatrix} 1 \\ 0 \\ 0 \\ -1 \end{bmatrix}$$
$$= \frac{1}{\sqrt{2}} \begin{bmatrix} 0 \\ 1 \\ -1 \\ 0 \end{bmatrix} = |\psi^-\rangle \quad (12)$$

And similar to the same analyses the following carriers will take the following operators to get the same results.

c) In case of the carrier Bell state $|\psi^+\rangle$:

| 2-bit element | U | $|\psi^o\rangle$ |
|---|---|---|
| 00 | X | $(X \otimes I)|\psi^+\rangle = |\phi^+\rangle$ |
| 01 | I | $(I \otimes I)|\psi^+\rangle = |\psi^+\rangle$ |
| 10 | iY | $(iY \otimes I)|\psi^+\rangle = |\phi^-\rangle$ |
| 11 | Z | $(Z \otimes I)|\psi^+\rangle = |\psi^-\rangle$ |

d) In case of the carrier Bell state $|\psi^-\rangle$:

| 2-bit element | U | $|\psi^o\rangle$ |
|---|---|---|
| 00 | XZ | $(XZ \otimes I)|\psi^-\rangle = |\phi^+\rangle$ |
| 01 | Z | $(Z \otimes I)|\psi^-\rangle = |\psi^+\rangle$ |
| 10 | iYZ | $(iYZ \otimes I)|\psi^-\rangle = |\phi^-\rangle$ |
| 11 | I | $(I \otimes I)|\psi^-\rangle = |\psi^-\rangle$ |

4) After $|\psi^o\rangle$ reached Bob, he starts to apply the appropriate unitary operations on the Bell states, measuring the two qubits and obtaining the 2-bit message element.

5) Bob performs the reverse operation of the encoding circuit, (decoding circuit) as:

$$(CNOT(H \otimes I)^{-1}) = (H \otimes I)CNOT \quad (13)$$

That is having the matrix representation:

$$B = \frac{1}{\sqrt{2}} \begin{bmatrix} 1 & 0 & 1 & 0 \\ 0 & 1 & 0 & 1 \\ 1 & 0 & -1 & 0 \\ 0 & 1 & 0 & -1 \end{bmatrix} \cdot \begin{bmatrix} 1 & 0 & 0 & 0 \\ 0 & 1 & 0 & 0 \\ 0 & 0 & 0 & 1 \\ 0 & 0 & 1 & 0 \end{bmatrix}$$
$$= \frac{1}{\sqrt{2}} \begin{bmatrix} 1 & 0 & 0 & 1 \\ 0 & 1 & 1 & 0 \\ 1 & 0 & 0 & -1 \\ 0 & 1 & -1 & 0 \end{bmatrix} \quad (14)$$

Therefore:

$$B|\phi^+\rangle = |00\rangle, \ B|\phi^-\rangle = |10\rangle, \ B|\psi^+\rangle = |01\rangle$$
$$B|\psi^-\rangle = |11\rangle \quad (15)$$

### 4.4 Comments

- In the implementation of the above protocol, it is essential that the two quantum channels used should be spatially separated all the way from Alice to Bob. This prevents an eavesdropper from accessing the two channels in one location and using the same procedure that should be used by Bob to get the original message.
- The following analysis illustrates the effectiveness of the protocol in counteracting the efforts of the eavesdropper.

All the Bell states used are pure maximally entangled states since if we consider one of them; $|\phi^+\rangle = \frac{|00\rangle + |11\rangle}{\sqrt{2}}$, then its density matrix $\rho$ is:

$$\rho = \frac{|00\rangle\langle 00| + |11\rangle\langle 00| + |00\rangle\langle 11| + |11\rangle\langle 11|}{2}$$

$$= \frac{1}{2}\begin{vmatrix} 1 & 0 & 0 & 1 \\ 0 & 0 & 0 & 0 \\ 0 & 0 & 0 & 0 \\ 1 & 0 & 0 & 1 \end{vmatrix} \quad (16)$$

Since $Tr(\rho^2) = 1$, then this is a pure state.
The partial trace over the first qubit is: $\rho^1 = I/2$
Since $Tr((\rho^1)^2) = \frac{1}{2}$ which is less than 1, then the first qubit is in a mixed state. Similarly, for the second qubit the same conclusion will be held. And as long as there is no unique mixed state for each separate quantum channel, then let assume each of them will be represented by one of these, called privileged mixed state. This could be obtained from the eigenvalues and eigenvectors of $\rho^1$ or $\rho^2$. The eigenvalues are equal to ½ and the eigenvectors are $|0\rangle$ and $|1\rangle$, so if we choose this specific case as the mixed state, then

$$\rho^1 \text{ or } \rho^2 = \frac{1}{2}|0\rangle\langle 0| + \frac{1}{2}|1\rangle\langle 1| \quad (17)$$

If the eavesdropper Eve has access to one quantum channel only, and makes a measurement she gets 0 or 1 with probability ½ for each case.

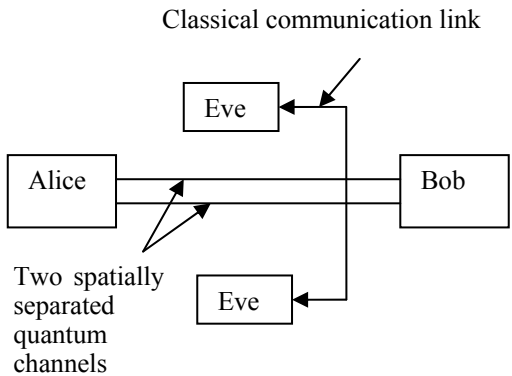

Figure 2: Synchronized attack with a classical link.

Let us, therefore assume that Eve can use the following attack which is rather difficult to implement. We call this attack "Synchronized attack together with a classical communication link", fig (3).
If Eve measure the qubit on quantum channel 1 she gets 0 with a probability ½, and assuming that a synchronized measurement is performed on quantum channel 2 she gets 0 with a probability ½. Therefore, for the Bell state $|\phi^+\rangle$, she gets 00 on the two quantum channels with probability ¼. Assuming that the Bell states used are uniformly distributed then each will have a probability of ¼. Then for message bits 00 Eve will get this result with probability 1/16. Similarly, for the other three combinations Eve will get the same results with the same probability. So if the message has length $N$, then the probability that Eve get the correct result is $\left(\frac{1}{16}\right)^{N/2} = \left(\frac{1}{4}\right)^N$.

## 5 EXAMPLES OF QUANTUM NOISE AND ITS EFFECT ON THE QUANTUM CHANNEL

In this section we examine some examples of quantum noise that could affect a quantum channel. These models are important in understanding the practical effects of the noise on quantum systems, and how noise can be controlled by techniques such as error-correction. Those models are bit flip; phase flip and both together (bit flip and phase flip). Of-course those models do not include all kinds of noise that could affect the quantum channel, there are others, but we chose these to analyze as they are more likely to occur. In our protocol we have two quantum channels, so the models of noise we mentioned above will be applied on both channels at random, i-e, we cannot know which model will affect which channel at a time, therefore we will study all different combinations of different models on the two channels, and then will analyze the last one in detail as it contains the greater combinations of the two other kinds.

In the first model (bit flip); the first qubit of Alice on the first channel after encoding the classical bits, could be flipped with probability (*p*), with the second qubit transmitted correctly. The second case is when the second qubit on the second channel, the half qubit of Bob, could be flipped with probability (*p*) where the first qubit transmitted correctly. The third case, if both qubits on both channels are flipped with probability ($p^2$).

In the second model (phase flip); also we have three cases as above, i-e, (anyone of the qubits will flip with probability (*p*), where the second will not), or the two qubits will flipped with probability ($p^2$).

In the third and last model (both bit and phase flip); all different combinations could happen; for example the first qubit could have a bit flip when the second qubit has a phase flip; or vice versa, and each of which will occur with probability $(p^2)$, etc. so we will introduce the analysis of this one as the most general one. Anyway, as we are using maximally entangled Bell states, all the models of quantum noise will just change the transmitted state to another one of the Bell states also, which makes it more confusing and harder to discover.

To protect the quantum state from the effect of the noise we would like to develop quantum error-correcting codes based upon similar idea of the classical error correcting codes. This idea is the repetition code, as that used by shor code (Nielsen, 2000).

In the following analysis we will consider the Bell state $|\phi^+\rangle = \frac{|00\rangle + |11\rangle}{\sqrt{2}}$ as an example since the other cases could be analyzed in a similar manner.

1. if the 1st bit has both bit flips and phase-flips and the 2nd bit remains as it is with
$$|\varphi^+\rangle \Rightarrow \frac{-|10\rangle - |01\rangle}{\sqrt{2}} = -|\psi^+\rangle \quad (18)$$
probability $(p^2)$ so

2. if the 2nd bit has both bit flips and phase-flips and the 1st bit remains as it is with probability (p2) so
$$|\varphi^+\rangle \Rightarrow \frac{-|01\rangle - |10\rangle}{\sqrt{2}} = -|\psi^+\rangle \quad (19)$$

3. if both bits have bit flips with only one of them has phase-flips with probability (p3) so
$$|\varphi^+\rangle \Rightarrow \frac{-|11\rangle - |00\rangle}{\sqrt{2}} = -|\varphi^+\rangle \quad (20)$$

4. if both bits have both bit flips and phase-flips and the with probability (p4) so
$$|\varphi^+\rangle \Rightarrow \frac{|11\rangle + |00\rangle}{\sqrt{2}} = |\varphi^+\rangle \quad (21)$$

5. if the 1st bit has bit flip and the 2nd bit has phase flip, with probability (p2) so
$$|\varphi^+\rangle \Rightarrow \frac{|10\rangle - |01\rangle}{\sqrt{2}} = -|\psi^-\rangle \quad (22)$$

6. if the 1st bit has phase flip and the 2nd bit has bit flip, with probability (p2) so
$$|\varphi^+\rangle \Rightarrow \frac{|01\rangle - |10\rangle}{\sqrt{2}} = |\psi^-\rangle \quad (23)$$

Note that the (–) sign in all the above relations introduced a global phase shift with no observable effect and could be dropped. So errors are introduced in 4 out of 6 cases above with appropriate probabilities. An error-correcting code scheme, like Steane code could then be used (Nielsen, 2000).

## 6 CONCLUSIONS

This paper introduced a new protocol for direct quantum communication making use of pure maximally entangled Bell states. Also, for efficiency purposes super dense coding is used, which is also based on entanglement, to double the transmission speed by sending two classical bits over one quantum channel. This protocol uses one step or one pass to end the message in a secure manner. It is essential that the two quantum channels used in the implementation be spatially separated all the way from Alice to Bob. To illustrate the security of the protocol, a hypothesized attack procedure used by Eve was considered that is called "synchronized attack together with a classical communication". Analysis was given to indicate that the probability of Eve getting the message is extremely small. Also, this type of attack is very difficult to implement. The effect of some quantum noise models was also considered, indicating the errors introduced. In this case some form of error-correcting procedure should be used.

## 7 FUTURE WORK

There are many aspects that could be considered to complete the above study. A few of them will be presented here:

1. Since the protocol is based on using pure maximally entangled Bell states, then it is essential to study procedure that could be used to get such states either from pure non entangled states, a process called concentration, or distillation and purification for mixed states.
2. It is essential to study entanglement degradation which depends on the length of the quantum channel. In particular,study of what is called Entanglement Sudden Death (ESD) phenomenon, which should be given appropriate attention, since it reduced sharply the distance over which entanglement is effective.
3. Other quantum noise effect should also be given due attention such as: depolarizing channel, amplitude damping, and phase damping.

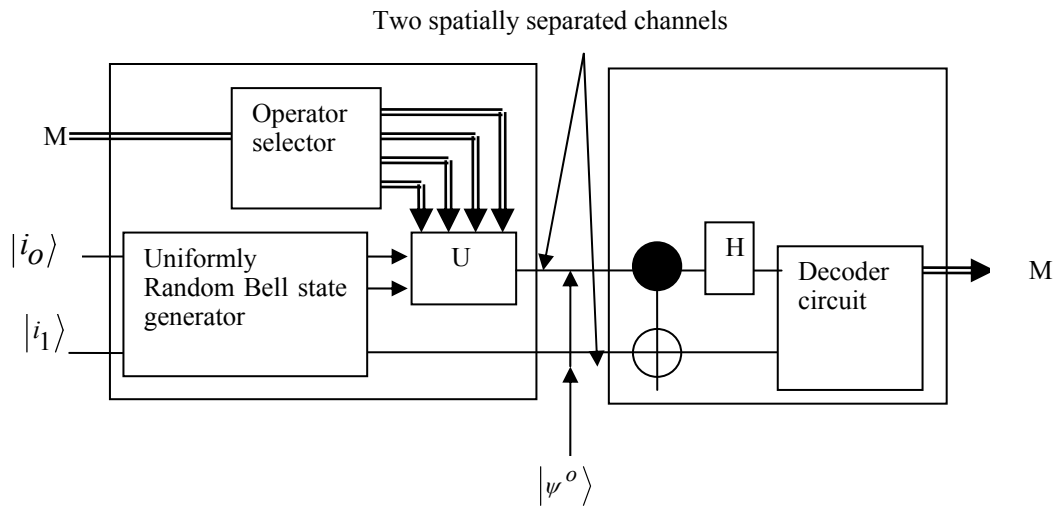

Figure 3: The block diagram of the coding and decoding circuit of the proposed protocol.